\documentclass{bmcart}

\usepackage{amsthm,amsmath}
\usepackage{amssymb}
\RequirePackage{hyperref}
\usepackage[utf8]{inputenc} 

\usepackage{graphicx} 
\usepackage[T1]{fontenc} 
\usepackage{listings} 
\usepackage{color} 

\definecolor{dkgreen}{rgb}{0,0.6,0}
\definecolor{gray}{rgb}{0.5,0.5,0.5}
\definecolor{mauve}{rgb}{0.58,0,0.82}

\startlocaldefs
\newcommand{\Circlearrowleft}{\circlearrowleft}  
\newcommand{\Circlearrowright}{\circlearrowright} 
\endlocaldefs

\begin{document}

\begin{frontmatter}

\begin{fmbox}
\dochead{Research}


\title{A model study of present-day Hall-effect circulators}


\author[
   addressref={aff1},                   
   corref={aff1},                       
   email={benedikt.placke@rwth-aachen.de}   
]{\inits{BP}\fnm{B.} \snm{Placke}}
\author[
   addressref={aff1, aff3},
   email={sfnbosco@gmail.com}
]{\inits{DD}\fnm{S. } \snm{Bosco}}
\author[
   addressref={aff1,aff2, aff3},
   email={d.divincenzo@fz-juelich.de}
]{\inits{DD}\fnm{D. P.} \snm{DiVincenzo}}


\address[id=aff1]{
  \orgname{Institute for Quantum Information, RWTH Achen University}, 
  \postcode{D-52056}                                
  \city{Aachen},                              
  \cny{Germany}                                    
}
\address[id=aff2]{%
  \orgname{Peter Grünberg Institute, Theoretical Nanoelectronics,
    Forschungszentrum Jülich},
  \postcode{D-52425}
  \city{J\"{u}lich},
  \cny{Germany}
}
\address[id=aff3]{%
  \orgname{J\"{u}lich-Aachen Research Alliance (JARA),
    Fundamentals of Future Information Technologiesh},
  \postcode{D-52425}
  \city{J\"{u}lich},
  \cny{Germany}
}


\begin{artnotes}
\end{artnotes}

\end{fmbox}



\begin{abstract} 
Stimulated by the recent implementation of a three-port Hall-effect microwave circulator of Mahoney {\em et al.} (MEA), we present model studies of the performance of this device.  Our calculations are based on the capacitive-coupling model of Viola and DiVincenzo (VD).  Based on conductance data from a typical Hall-bar device obtained from a two-dimensional electron gas (2DEG) in a magnetic field, we numerically solve the coupled field-circuit equations to calculate the expected performance of the circulator, as determined by the $S$ parameters of the device when coupled to 50$\Omega$ ports, as a function of frequency and magnetic field.  Above magnetic fields of 1.5T, for which a typical 2DEG enters the quantum Hall regime (corresponding to a Landau-level filling fraction $\nu$ of 20), the Hall angle $\theta_H=\tan^{-1}\sigma_{xy}/\sigma_{xx}$ always remains close to $90^\circ$, and the $S$ parameters are close to the analytic predictions of VD for $\theta_H=\pi/2$.  As anticipated by VD, MEA find the device to have rather high (k$\Omega$) impedance, and thus to be extremely mismatched to $50\Omega$, requiring the use of impedance matching.  We incorporate the lumped matching circuits of MEA in our modeling and confirm that they can produce excellent circulation, although confined to a very small bandwidth.  We predict that this bandwidth is significantly improved by working at lower magnetic field when the Landau index is high, e.g. $\nu=20$, and the impedance mismatch is correspondingly less extreme.  Our modeling also confirms the observation of MEA that parasitic port-to-port capacitance can produce very interesting countercirculation effects. 
\end{abstract}




%

\end{frontmatter}


\section{Introduction}

The circulator is an important component of modern low-temperature microwave engineering.
It is essential for several applications, including the measurement and control of solid state qubits \cite{DiCarlo, 11qubits} to thermal noise exclusion \cite{DiCarlo}.
An ideal circulator is a lossless non-reciprocal three-port device that cyclically routes a signal to the next port. Its behavior is captured by the scattering ($S$) matrices \cite{Pozar} 
\begin{flalign}
&&
    \label{eq:S-circulator}
    S_{\Circlearrowright} =
    \begin{pmatrix}
        0 & 0 & 1 \\
        1 & 0 & 0 \\
        0 & 1 & 0
    \end{pmatrix}
    &&
    \mathrm{and}
    &&
        S_{\Circlearrowleft} =
    \begin{pmatrix}
        0 & 1 & 0 \\
        0 & 0 & 1 \\
        1 & 0 & 0
    \end{pmatrix},
    &&
\end{flalign}
where the arrows indicate the direction of circulation.  Both senses of rotation will be relevant in the present work.
The conventional way of implementing passive circulators exploits the classical Faraday effect \cite{Pozar, Hogan1, Hogan2}. Although these devices can be very efficient in terms of loss, in the microwave regime they are quite bulky, with a physical scale on the order of centimeters. In fact, this construction has a fundamental scalability issue: since it relies on a wave-interference phenomenon, the minimum size is set by the wavelength at which they operate.

Recently, Viola and DiVincenzo (VD) proposed an alternative implementation for a passive circulator, which points the way towards better scalability \cite{Viola-DiVincenzo}. The main message of their work is that non-reciprocal electrical conduction can lead to well-performing non-reciprocal devices. In particular, ideal circulation at specific frequencies is achieved by reactively coupling electrodes to the edges of a two-dimensional conductor such as a two-dimensional electron gas (2DEG) with a strong Hall effect.  By the construction of Carlin \cite{Viola-DiVincenzo, Carlin, Carlin1}, a 2DEG with three capacitively coupled edge terminals becomes a three-port circulator when the common grounds of the three ports are kept out of contact with the Hall conductor (i.e., the 2DEG should float with respect to the port ground).

The model proposed by VD applies for any values of Hall conductance, requiring the solution of coupled field-circuit equations.  However, the results presented by VD were limited to the case when the Hall angle $\theta_H\equiv\tan^{-1}\sigma_{xy}/\sigma_{xx}$ is exactly at its maximal value $\pi/2$, corresponding to parameter values (magnetic field, carrier density) for which the material conductance is on a quantum Hall plateau and the Landau level filling parameter $\nu$ is an integer.  Under these conditions the VD equations are solvable in closed form; for the more general case considered here, a numerical solution is needed. These solutions, and their consequences for the device $S$ parameters, will be presented below.

An experimental realization of a Hall effect microwave circulator has now been reported by Mahoney {\em et al.} (MEA) \cite{Reilly}. MEA observes that impedance matching and parasitic capacitive coupling between neighboring ports play a key role for the behavior of their circulator. Impedance matching is essential to achieve good circulation, and the appropriate level of parasitic capacitance has the surprising consequence of inverting the direction of circulation in going from one magnetic field or frequency to another. By engineering the coupling between electrodes, this interesting property might be exploited for novel applications, where tiny changes of field could reverse the direction of circulation. 

The present work is developed in light of MEA's results.  First, we employ the VD model, solving the field problem of a three-port Carlin circulator with realistic values of Hall angle obtained from typical 2DEG magnetoconductance characteristics. These characteristics involve a low magnetic-field regime in which small oscillations in the diagonal and off-diagonal conductances are observed (the Shubnikov–de Haas regime); as these oscillations grow to near 100\% amplitude, one enters the quantum Hall regime in which plateaus occur in the off diagonal ($\sigma_{xy}$) conductance, and the Hall angle stays very near $\pi/2$, oscillating slightly away from this value as the conductance passes from plateau to plateau.  We see that our numerical calculations of the device performance are to a first approximation given by the analytic results of VD throughout the quantum Hall regime, but with noticeable departures from ideal behavior.
We also incorporate in our calculations of the device response a complete circuit description including the impedance matching circuit and parasitic effects as reported by MEA. We find that impedance matching is essential for the realization of circulation; without matching the reflection coefficient of the device is very close to one, making the response very far from the desired behavior.  We note, however, the impedance matching limits the versatility of the device: matching can be effectively performed only for specific combinations of parameters, i.e., for specific frequencies or magnetic fields.

Our paper is organized as follows.
In Sec. \ref{sec:basic-analysis}, we  compute the $S$ parameters of an ideal Carlin circulator, when the device and the external circuit have equal impedance.
In Sec. \ref{sec:impedance-matching}, we incorporate impedance matching circuits optimized for two specific magnetic fields and we study the response in the two cases.
In Sec. \ref{sec:parasitics}, we include the parasitic coupling between electrodes: we confirm the magnetic field dependent change in the direction of circulation observed in \cite{Reilly} and we quantitatively analyze this phenomenon.

\section{Basic Analysis \label{sec:basic-analysis}}

The geometry considered throughout this work is shown in Fig \ref{fig:hall_bar}. 
We consider a two-dimensional material lying in the $(x,y)$-plane subject to a uniform perpendicular magnetic field $B$. Three electrodes of equal length $L$ are capacitively coupled to the material and  they are symmetrically distributed around its boundary, parametrized by coordinate $s$. This three-electrode device forms a circulator of the "first Carlin" type \cite{Viola-DiVincenzo}. To model this circulator, we follow VD \cite{Viola-DiVincenzo}; we assume that the capacitors are much longer than the gaps between them, also corresponding to the geometry of MEA.  The VD model neglects any capacitances except for the external electrode capacitors, which should be reasonable for the aspect ratios used here.  The work of VD also indicates that for Hall angle $\theta_H\lesssim\pi/2$, the device response should be sensitive only to the edge dimensions and not to the shape of the conductor.  Thus, we do not expect any large differences between the square device as modelled here and the circular device of MEA.
 
As a first step, we calculate the electric potential $V(x,y)$ inside the 2D Hall conductor. $V(x,y)$ satisfies the 2-dimensional Laplace equation \cite{Wick}
\begin{equation}
\nabla^2V(x,y)=0,
\label{lapl}
\end{equation}
with the boundary conditions that we now describe.

The boundary conditions must account for the capacitive coupling to external electrodes and for the deflection of electrons due to the  applied magnetic field. In the frequency domain, they can be stated as \cite{Viola-DiVincenzo}, for excitations at frequency $\omega$:
\begin{equation}
    -\sigma \hat{n}_H \cdot\vec{\nabla}V(s, \omega)
    = i \omega c(s)\left( \overline{V}(\omega) - V(s, \omega) \right),
    \label{eq:capacitive-boundary}
\end{equation}
where $\sigma$ is the magnitude of the Hall conductivity tensor (assumed frequency independent), $\hat{n}_H$ is the unit vector rotated by $\theta_H$ with respect to the direction normal to the boundary and $\overline{V}(\omega)$ is the Fourier transform of the time-dependent voltages applied at the electrodes.  It is assumed that $\omega$ is low enough, and the device is small enough that retardation effects are negligible, assuring the applicability of Eq. (\ref{lapl}). 
The phenomenological function $c(s)$ models the local coupling with the electrodes; it has the dimensions of capacitance per unit length. In our treatment, it takes the form
\begin{flalign}
&&
c(s)=\left\{
\begin{array}{ll}
      c, & \lvert s-s_i-L/2 \rvert \leq L/2 \\
      0, & \lvert s-s_i-L/2 \rvert > L/2 
\end{array} 
\right. 
,
&&
\textrm{with}
&&
i=1,2,3.
&&
\end{flalign}
$s_i$ is the position along the perimeter of the starting point of lead $i$.
The parameter $c$ can be estimated for certain geometries from theory, e.g. \cite{Volkov, Mikhailov, Glazman}, or directly extracted from experiment. It has contributions from both classical electrostatic coupling and from quantum capacitance due to screening effects \cite{Vignale, Buttiker, Buttiker2}. 
A rule of thumb is that quantum capacitance starts to play an important role when the distance between electrodes and quantum Hall droplet is comparable with the electron screening length.
In MEA \cite{Reilly}, this condition is apparently not met, and we believe that the coupling is strongly dominated by the classical geometric capacitance; this is no problem for the realization of a circulator.

Once the electric potential is found, all the relevant quantities are straightforwardly computed. 
In particular, the current at the $i$th electrode is related to the boundary potential by \cite{Viola-DiVincenzo}
\begin{equation}
    I_i(\omega) = - \sigma \int_{s_i}^{s_i+L} ds \hat n_H \cdot \vec{\nabla} V(s, \omega).
    \label{eq:current-rotated}
\end{equation}
From this relation, we readily obtain the admittance matrix of the device $Y_c$ by applying harmonic drives at each electrode separately and using the superposition principle of linear circuit theory.

Finally, we convert the admittance matrix into a scattering ($S$) matrix using the relation \cite{Pozar}
\begin{equation}
    S = (\mathcal{I} - Z_0 Y)(\mathcal{I} + Z_0 Y)^{-1},
    \label{eq:SfromY}
\end{equation}
with $\mathcal{I}$ being the identity matrix and $Z_0$ being the characteristic impedance of the external circuit.
Exploiting the unitarity of $S$ and comparing with Eq. (\ref{eq:S-circulator}), one can establish a criterion  \cite{Viola-DiVincenzo} to quantify the circulating behavior of the device:
\begin{subequations}
\label{eq:Q_cw_acw}
\begin{flalign}
    Q_{\Circlearrowleft} \equiv \lvert S_{12} \rvert + \lvert S_{23}\rvert + \lvert S_{31}\rvert\leq 3, \\
    Q_{\Circlearrowright} \equiv \lvert S_{21}\rvert + \lvert S_{32}\rvert + \lvert S_{13}\rvert\leq 3,
\end{flalign}
\end{subequations} 
where the equality corresponds to perfect circulation in the direction of the arrow. 
When the equality is not met, a fraction of the signal is reflected back or circulates in the opposite direction, and the device partially looses its chirality. To quantify this effect, we show in Fig. \ref{fig:scatter-plot} a scatter plot that relates the parameters $ Q_{\Circlearrowleft}$ and $ Q_{\Circlearrowright}$ obtained from a set of random unitary $S$-matrices. From the Figure, it appears that even tiny variations of  $Q$  from 3 can lead to a significant backward circulation and consequently to a noticeable loss of chirality, e.g. for $Q_{\Circlearrowleft}= 2.8$, $Q_{\Circlearrowright}$ can attain a value as large as 1. 

Taking the limit $\theta_H\rightarrow\pi/2$ in Eq. (\ref{eq:capacitive-boundary}), the equation for the potential along the perimeter decouples from the bulk. VD found an analytic solution for this boundary problem \cite{Viola-DiVincenzo} and they showed that the device in Fig. \ref{fig:hall_bar} then behaves as an ideal circulator at frequencies
\begin{equation}
    \label{eq:circulator-f-Viola}
    \omega_{c} = \pi \frac{\sigma}{c L}(1 + 2n), n \in \mathbb{N}_0,
\end{equation}
when the voltage of the three electrode is referenced to a common ground and when the device is perfectly matched, i.e. the external circuit equals the impedance of the circulator $Z_0 = R_c\equiv(2 \sigma)^{-1}$.

The $\theta_H<\pi/2$ case is difficult to deal with analytically as the boundary potential couples with the bulk potential. Therefore, we have computed a finite-difference numerical solution for the Laplace equation. In all the following simulations, we used $c L=50$fF, which agrees with estimated parameters in the recent experiment of MEA \cite{Reilly}, $L=1\mathrm{mm}$ and the spacing between the electrodes $sp=0.33\mathrm{mm}$. Moreover, we extracted $\sigma$ and $\theta_H$ from the experimental data shown in Fig. \ref{fig:quantum-hall-effect-real}; this data, which is entirely generic for heterostructure 2DEGs, is taken from a device used in the advanced (masters) physics lab course of the second physics institute of Aachen University.\footnote{\tiny  \mbox{https://institut2a.physik.rwth-aachen.de/de/teaching/praktikum/Anleitungen/QTV\_instructions.pdf}}
Figure \ref{fig:S-circ-ideal-match} shows the $Q$-parameters as a function of frequency and magnetic field for a perfectly matched device (note that perfect matching requires a magnetic field dependent $Z_0$). From the plot, we confirm that at the frequency in Eq. (\ref{eq:circulator-f-Viola}) almost perfect circulation in the anticlockwise direction can be achieved.
Comparing with Figure \ref{fig:quantum-hall-effect-real}, we notice that the highest values of $Q$ are obtained at magnetic fields corresponding to the quantum Hall plateau, but good circulating performances are guaranteed also in the transition regions and in the Shubnikov-de Haas regime. In particular, we observe that in the latter regime the device has a greater bandwidth.


The perfectly matched circulator described here is unfortunately a purely theoretical device. In the quantum Hall regime, the impedance of the device is
\begin{equation}
R_c=\frac{2h}{e^2\nu} \approx \frac{50\mathrm{k\Omega}}{\nu},
\label{eq:conductance-quantum-hall}
\end{equation}
with $\nu$ being the filling factor \cite{QuantumHallGirvin}, far greater than the characteristic impedance of standard microwave circuits $Z_0=50\Omega$. Thus, use of impedance matching techniques is essential.

\section{Augmented network for impedance matching \label{sec:impedance-matching}}

To have a more realistic picture of the performance of the device, we include lumped-element impedance matching circuits as suggested by MEA.
We focus our analysis on the augmented network shown in Fig. \ref{fig:circ-impedance-matching-sketch}, as proposed in \cite{Reilly}.
Simple $LC$-circuits at each port are used to match the impedance of the device with the external circuit. This network, like any impedance-matching circuit, has the drawback of working only at specific frequencies $\omega_m$, limiting the versatility of the circulator. Fig. \ref{fig:circ-impedance-matching-sketch} also shows parasitic capacitances coupling each pair of neighboring electrodes, as suggested by MEA.  We will neglect these for the time being, returning to their analysis in the following section.
From standard circuit theory (see Chap. 3 of \cite{Newcomb}), one can write the admittance matrix of the augmented device
\begin{equation}
    Y_{a} = \left(
    i\omega L_m \mathcal{I}+ \left(i \omega C_m \mathcal{I} + Y_c \right)^{-1} \right)^{-1}.
    \label{eq:impedance-matrix-full2}
\end{equation}

The values of $L_m$ and $C_m$ are fixed by the standard design formulas (Chap. 5.1 of \cite{Pozar})
\begin{subequations}
\label{eq:impedance-matching-real}
\begin{align}
    \label{eq:impedance-matching-real-B}
    C_m\omega_m &=  \frac{1}{Z_0} \sqrt{1 - \frac{Z_0}{R_c}} ,\\
    \label{eq:impedance-matching-real-X}
    L_m\omega_m &= Z_0 \sqrt{1 - \frac{Z_0}{R_c}},
\end{align}
\end{subequations}
We will take $Z_0=50\Omega$ as the characteristic impedance of the standard transmission system.
According to equation (\ref{eq:conductance-quantum-hall}), the impedance of the circulator $R_c$ depends on the filling factor $\nu$. Hence, the impedance matching circuit works only for specific filling factors, setting an additional constraint on the regime of parameters which guarantees good circulation.

Figure \ref{fig:S-circ-50-ohms-match} shows a plot of the $Q$-parameters when the matching is optimized for filling factor $\nu=8$ and for the first circulation frequency, i.e. $\omega_m=\omega_c(n=1)$. As expected, the device behaves as the theoretical perfectly matched device, in Fig. \ref{fig:S-circ-ideal-match}, only near the filling factor and frequency at which the matching is performed. 
For the parameters needed for good impedance matching, the effective bandwidth of the device is limited, as expected given the very large mismatch to be overcome.

Since the bandwidth of the impedance matching circuits increases if $Z_0/R_c$ can be decreased, the bandwidth of the circulator can be improved by matching the device at higher filling factor.
Figure \ref{fig:S-circ-50-ohms-match-nu20} shows the performance of a device matched at $\nu=20$. The bandwidth of this device is clearly greater than in Fig. \ref{fig:S-circ-50-ohms-match}. However, as we approach the Shubnikov-de Haas regime at low magnetic field, where the magnetoconductance oscillations become weak and the Hall angle decreases, the maximum value of the $Q$-parameter decreases.
Hence, when engineering the device there is a trade-off between bandwidth and circulation performance to be accounted for.

\section{Effect of parasitic capacitances \label{sec:parasitics}}

We now extend the analysis to include the effect of the parasitic capacitances of Fig. \ref{fig:circ-impedance-matching-sketch}.  We assume three equal capacitances with value $C_p$.  The augmented-network analysis of the new admittance matrix gives the formula

\begin{equation}
    Y_{a} = \left(
    i\omega L_m \mathcal{I}+ \left(i \omega C_m \mathcal{I}+ i \omega C_p (2\mathcal{I}-S_{\Circlearrowleft}-S_{\Circlearrowright}) + Y_c \right)^{-1} \right)^{-1}.
    \label{eq:impedance-matrix-full}
\end{equation}

We will consider the $\nu=8$ case, examining different values of $C_p$.
Figure \ref{fig:S-circ-weak-parasitics} shows the results when $C_p=2\mathrm{fF}$. This parasitic component is evidently very small and its effect is negligible, as one can see comparing with Fig. \ref{fig:S-circ-50-ohms-match}.
Qualitatively, the parasitic capacitances start to influence the device response when $1/R_c$ and $\omega C_p$ are of the same order of magnitude, since then the parasitic and the direct channels carry currents of equal magnitude, which can then interfere with one another. In our case, this condition corresponds to $ C_p \approx 2\sigma/\omega\approx 20\mathrm{fF}$ for the first circulation frequency. Figure \ref{fig:S-circ-parasitics} shows $Q$-parameters for the realistic parasitic coupling value $C_p=120\mathrm{fF}$. The surprising effect of the additional capacitive channels is to introduce significant circulation in the reverse direction, as experimentally observed in \cite{Reilly}. The direction of circulation becomes magnetic-field dependent and is seen to change in the vicinity of the curve defined by $\omega=\omega_c(n=1)$. Interestingly, for the parameters chosen, both circulators behave almost equally well, with a maximum of $Q_{\Circlearrowleft}^{max}=2.75$ and $Q_{\Circlearrowright}^{max}=2.88$. 
Finally, for significantly larger values of the parasitic capacitances, they dominate the device response and circulation in both directions is suppressed, as shown in Fig. \ref{fig:S-circ-strong-parasitics}.

To gain more insight into the reverse circulation phenomenon, we investigate the dependence of the $Q$-parameters on $C_p$. Figure \ref{fig:Q-of-Cp} shows the maximum value of $Q$ for the two directions of circulation as a function of the parasitic coupling.
Strikingly, reverse circulation can be more effective (larger Q) than the direct one, and it degrades more slowly as $C_p$ increases. 
Finally, we observe that for certain coupling values, e.g. $C_p\approx40\mathrm{fF}$, almost perfect circulation can be achieved in both directions (at different frequencies, of course). This phenomenon might be exploited for novel applications, where tiny changes of magnetic field can reverse the direction of circulation.

\section{Conclusion}
We have investigated an implementation of a quantum Hall effect circulator.
We proved that realistic variations of Hall angle occurring once the quantum Hall regime has been entered (at around 1.5T for the 2DEG considered)  do not alter significantly the fundamental behavior of the device investigated in \cite{Viola-DiVincenzo}, even for a large Landau-level index ($\nu=20$).
Moreover, we addressed two important microwave-engineering issues, impedance matching and parasitic capacitances.
Although the presence of an impedance matching circuit decreases the versatility of the device, we proved that the performance of the circulator can be optimized either to have a large bandwidth or to operate in a large range of magnetic field.
Finally, analyzing the effect of parasitic capacitances, we observed a reverse circulation phenomenon, expected from recent experiments \cite{Reilly}. Interestingly, we found that depending on the coupling strength the reverse circulation is comparable and for certain parameters even better than the direct one, opening up to new possible applications.


\begin{backmatter}

\section*{Competing interests}
  The authors declare that they have no competing interests.


\section*{Acknowledgements}
  The authors would like to thank D. Reilly, A. Mahoney, A.C. Doherty, and T. Leonhardt for useful discussions.
This work was supported by the Alexander von Humboldt foundation.


\bibliographystyle{bmc-mathphys} 
\bibliography{lit}      



\pagebreak
\section*{Figures}

\begin{figure}[h!]
\includegraphics[scale=0.7]{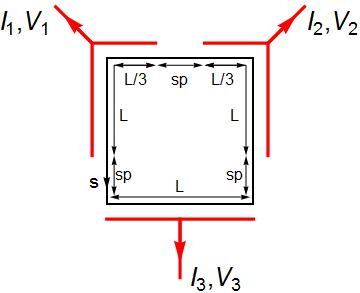}
\caption{Three-terminal capacitively coupled Hall bar. Three metal electrodes of length $L=1$mm are capacitively coupled to a square Hall bar. The electrodes are symmetrically placed around the perimeter of the device, parametrized by $s$, and they are separated by a gap $sp=L/3=0.33$mm.
The convention of currents and voltages is shown in the plot.}
\label{fig:hall_bar}
\end{figure}

\begin{figure}[h!]
    \includegraphics[width=0.7\textwidth]{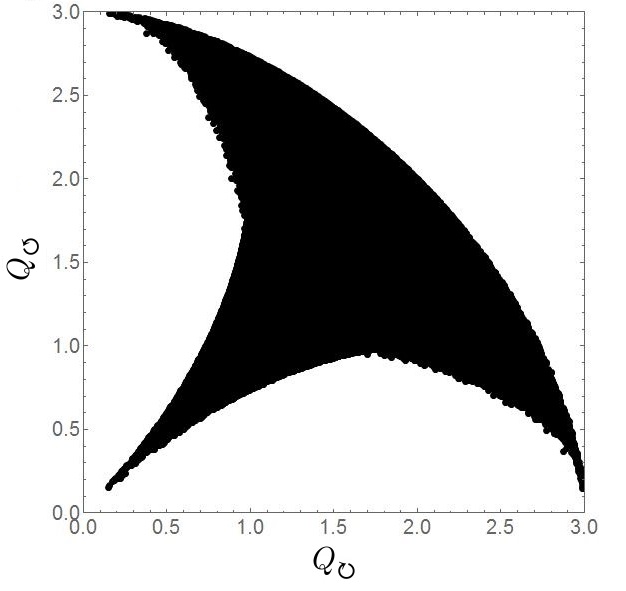}
    \caption{Relation between $Q_{\circlearrowleft}$ and $Q_{\circlearrowright}$.
    The scatter plot was generated by computing the two  $Q$-parameters of a set of random unitary $S$-matrices. The solutions at the corners $(0,0)$, $(0,3)$ and $(3,0)$ exist (for example, they are obtained respectively from $\mathcal{I}$, $S_{\circlearrowleft}$ and $S_{\circlearrowright}$); they do not appear in the plot as they are statistically not likely to occur in a set of random $S$ matrices. In addition, the point $(2,2)$ is the well-known upper bound for reciprocal 3x3 $S$-matrices, see, e.g. Sec. 4.9 of \cite{Carlin}. }
    \label{fig:scatter-plot}
\end{figure}

\begin{figure}[h!]
    \includegraphics[width=0.8\textwidth]{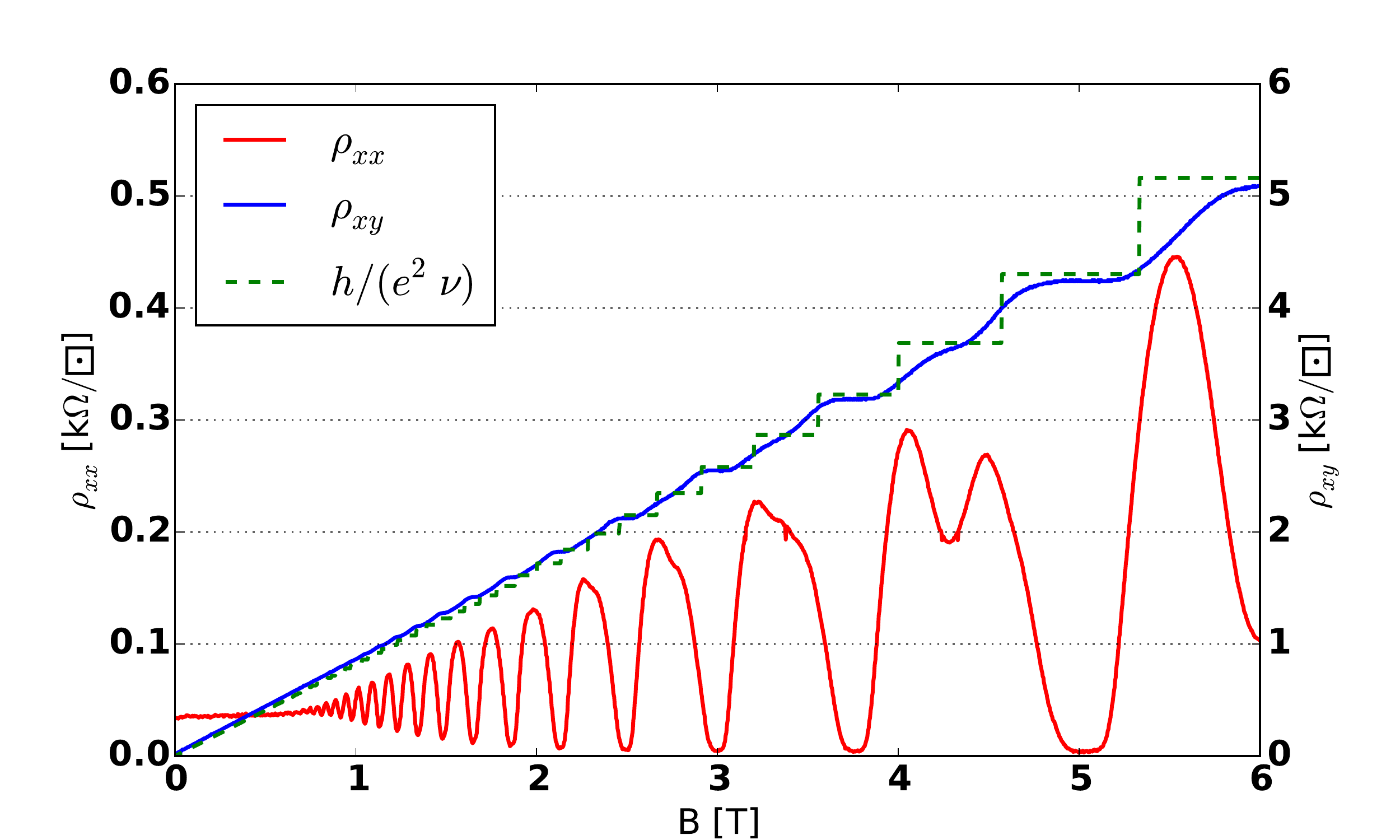}
    \caption{Magnetic field dependence of the components of the resistivity tensor for a typical 2DEG heterostructure (InGaAs/InP) at $T=4$K. For magnetic fields greater than 1.5T, the resistivity develops plateau regions characteristic of the quantum Hall regime, where $\rho_{xy}$ is constant and the diagonal element $\rho_{xx}$ goes to zero.
     This data shows negligible evidence of the fractional quantum Hall effect. }
    \label{fig:quantum-hall-effect-real}
\end{figure}

\begin{figure}[h!]
    \centering{}
    \includegraphics[width=\textwidth]{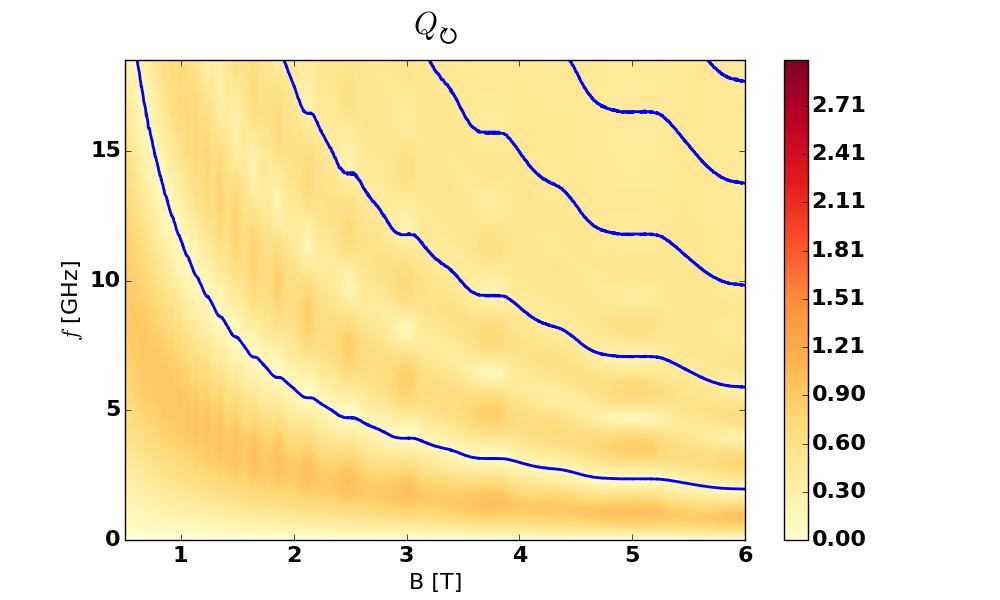} \\
    \includegraphics[width=\textwidth]{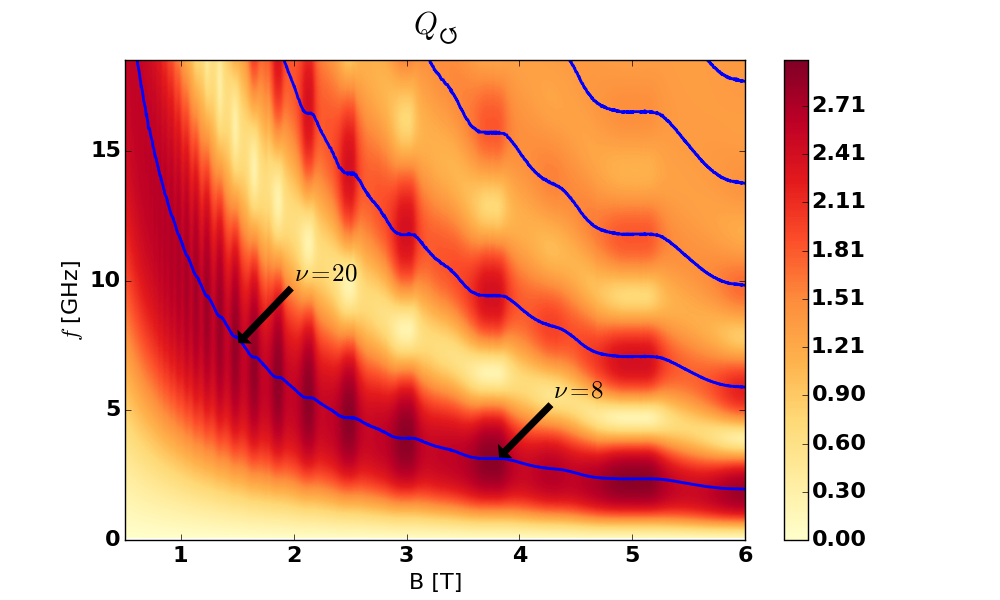}
    \caption{ $Q$-parameters of a perfectly matched three terminal circulator. The device geometry is shown in Fig. \ref{fig:hall_bar}, and we use a capacitance $cL=50$fF; the magnitude of the conductivity tensor $\sigma=\sqrt{\sigma_{xx}^2+\sigma_{xy}^2}$  and the Hall angle $\theta_H$ are extracted from the experimental data in Fig. \ref{fig:quantum-hall-effect-real}.
Along the blue lines, corresponding to Eq. (\ref{eq:circulator-f-Viola}), the value $Q_{\circlearrowleft}=3$ is exactly attained in the limit $\theta_H=\pi/2$, giving perfect circulation in the anticlockwise direction.
We indicate the plateaus corresponding to filling factors $\nu=20$ and $\nu=8$, where impedance matching will be performed. }
    \label{fig:S-circ-ideal-match}
\end{figure}

\begin{figure}[h!]
    \centering{}
    \includegraphics[width=.9\textwidth]{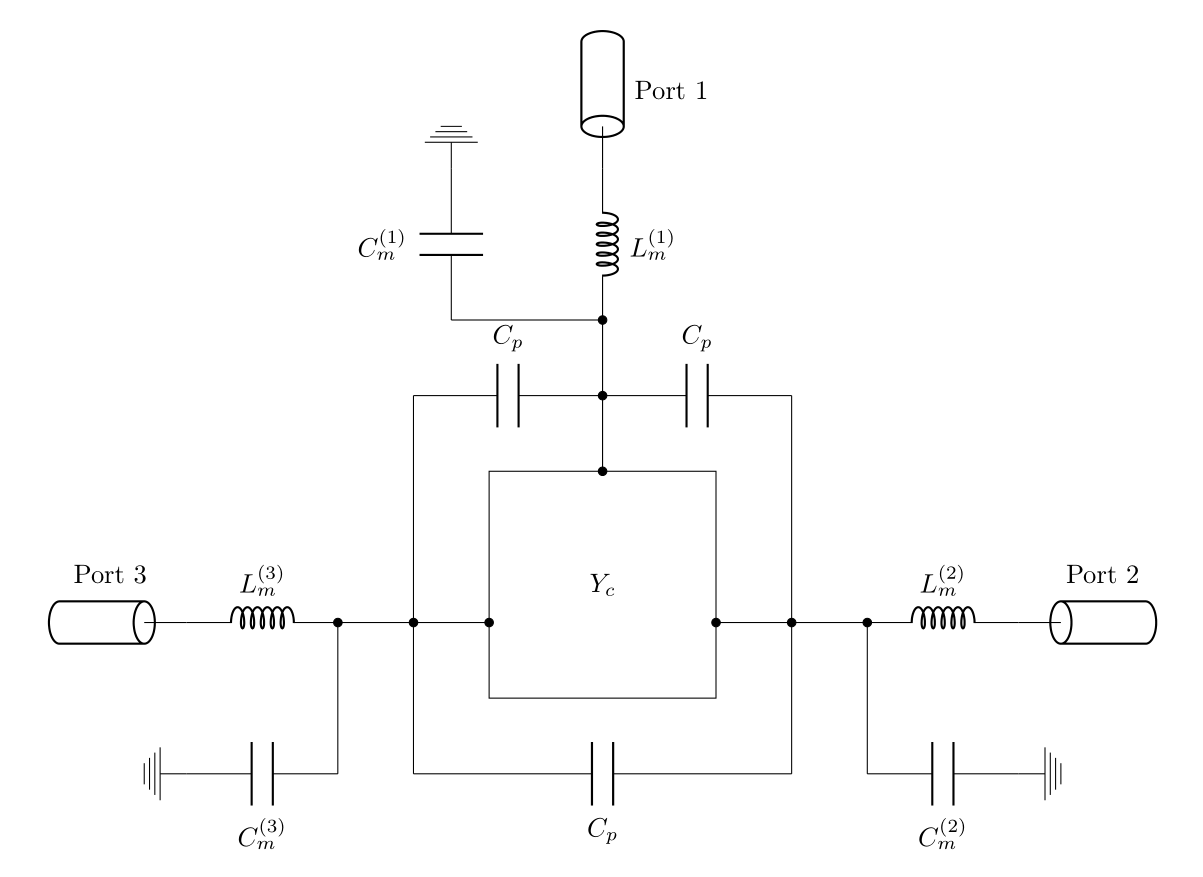}
    \caption{Augmented network used for impedance matching and including the parasitic capacitances that couple the three terminals. 
The impedance of each port is separately matched to the one of the circulator by $LC$-circuits, as proposed in \cite{Reilly}.}
    \label{fig:circ-impedance-matching-sketch}
\end{figure}

\begin{figure}[h!]
    \centering{}
    \includegraphics[width=\textwidth]{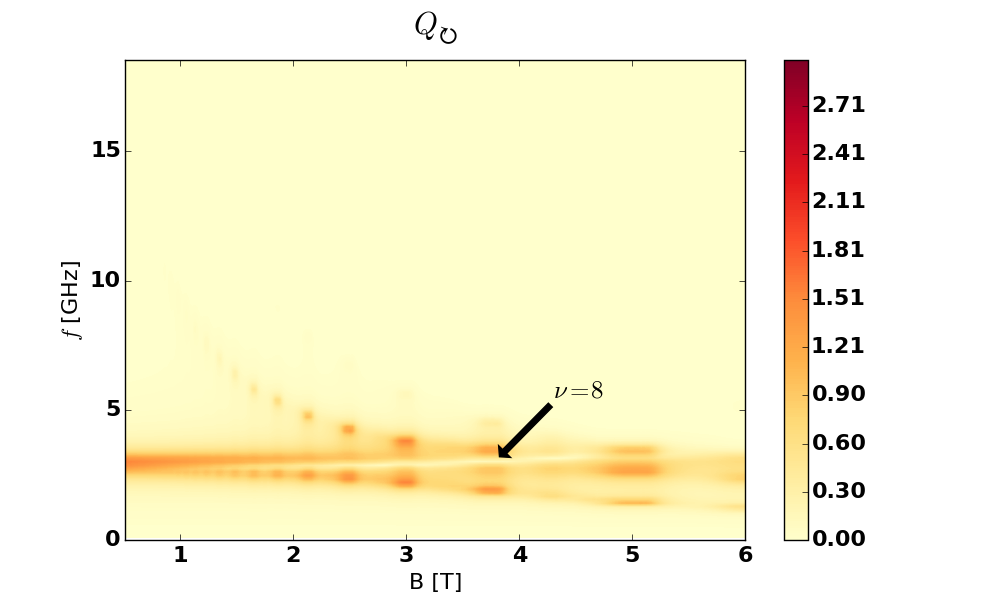} \\
    \includegraphics[width=\textwidth]{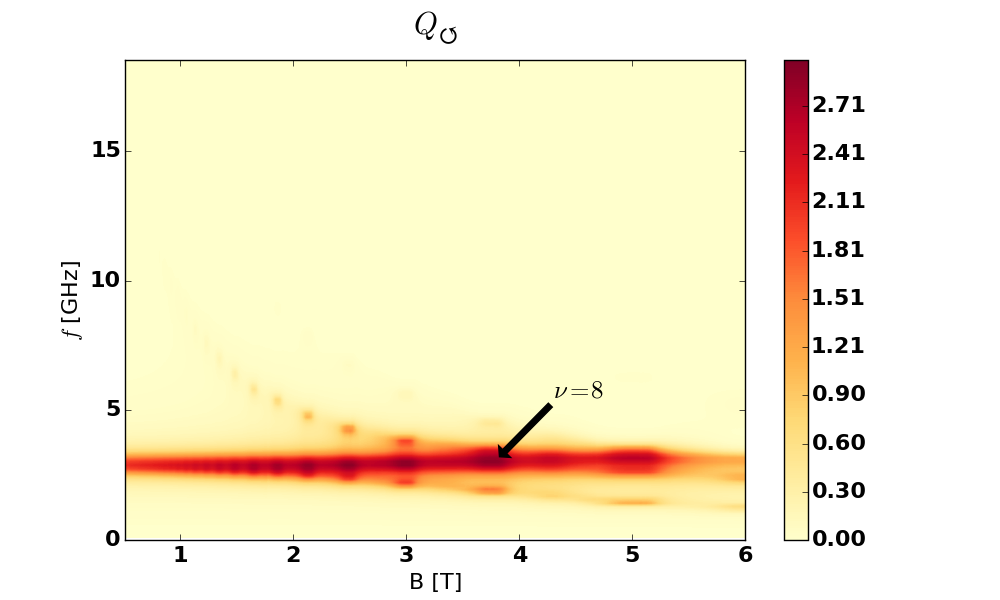}
    \caption{$Q$-parameters of a three-port circulator, matched to the external impedance of $50\Omega$. The matching is optimized for filling factor $\nu = 8$; here, we neglect the parasitic coupling between electrodes.
 Comparing with Fig. \ref{fig:S-circ-ideal-match}, good circulation is achieved only at the magnetic fields corresponding to $\nu=8$, and the bandwidth of the device shrinks notably due to the high impedance mismatch to be overcome.}
    \label{fig:S-circ-50-ohms-match}
\end{figure}

\begin{figure}[h!]
    \centering{}
    \includegraphics[width=\textwidth]{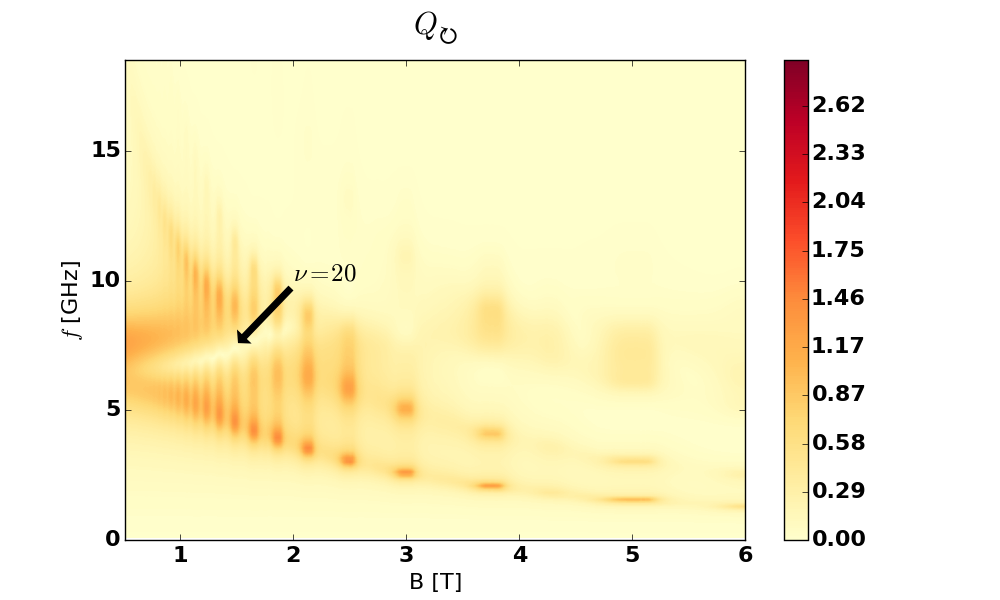} \\
    \includegraphics[width=\textwidth]{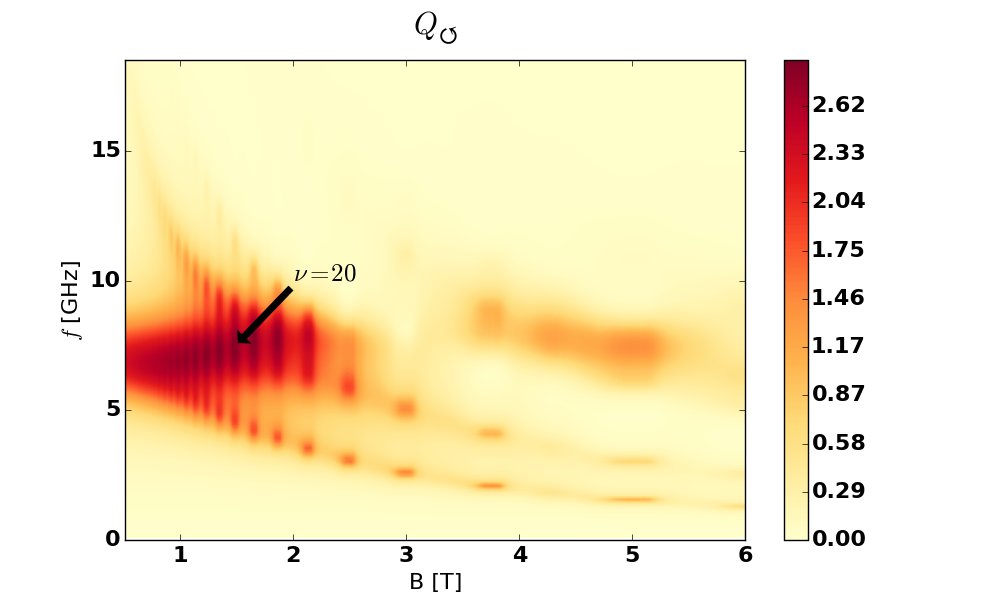}
    \caption{$Q$-parameters of a three-port circulator, matched to the external impedance of $50\Omega$. The matching is optimized for filling factor $\nu = 20$; here, we neglect the parasitic coupling between electrodes.
     Comparing with Fig. \ref{fig:S-circ-50-ohms-match}, the bandwidth is greater, but the circulation performance decreases as the Shubnikov-de Haas regime is approached.}
    \label{fig:S-circ-50-ohms-match-nu20}
\end{figure}

\begin{figure}[h!]
    \centering{}
    \includegraphics[width=\textwidth]{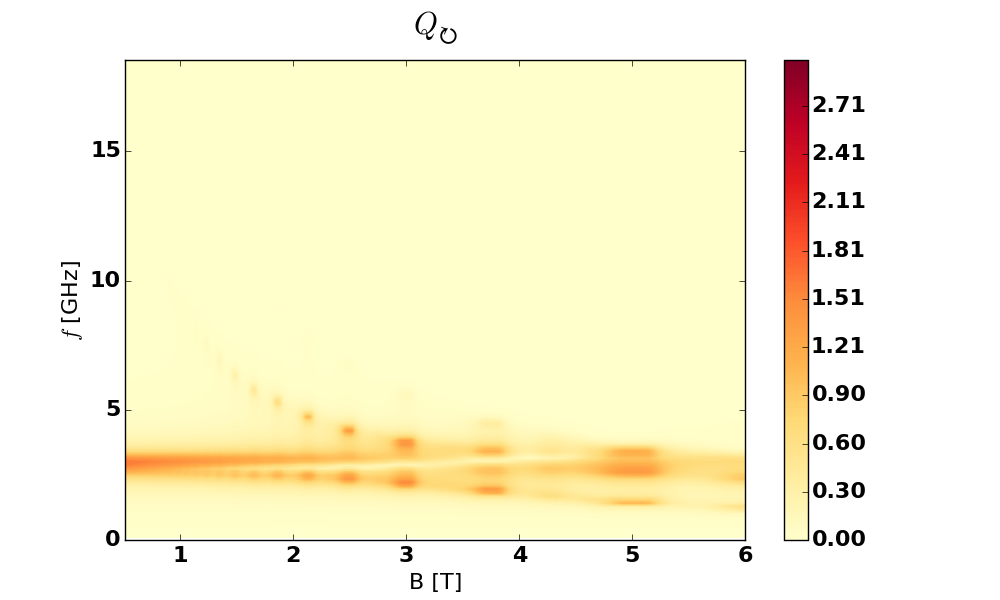} \\
    \includegraphics[width=\textwidth]{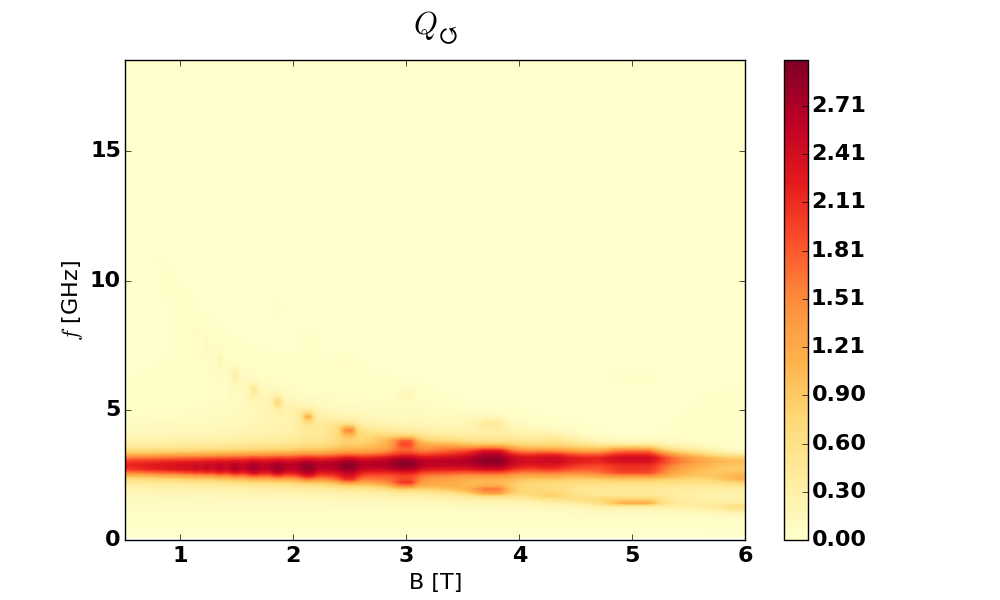}
    \caption{Effects of parasitics on the $Q$-parameters of a three-port circulator. The circulator is matched to the external impedance of $50\Omega$ and the matching is optimized for filling factor $\nu = 8$.
    We used here $C_p=2$fF. As the parasitic capacitances are very small, their effect on the response is negligible, see Fig. \ref{fig:S-circ-50-ohms-match}.}
    \label{fig:S-circ-weak-parasitics}
\end{figure}

\begin{figure}[h!]
    \centering{}
    \includegraphics[width=\textwidth]{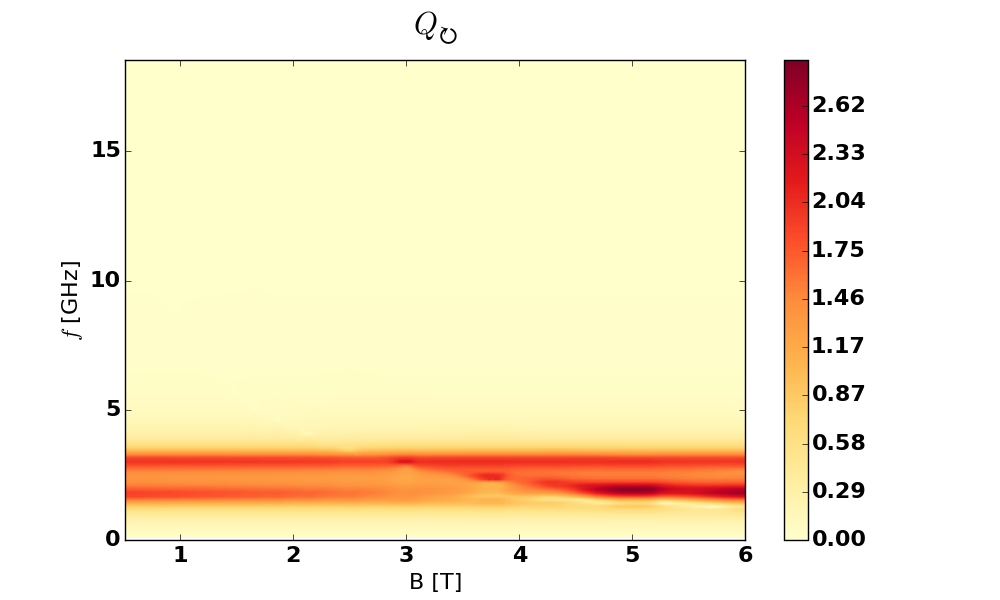} \\
    \includegraphics[width=\textwidth]{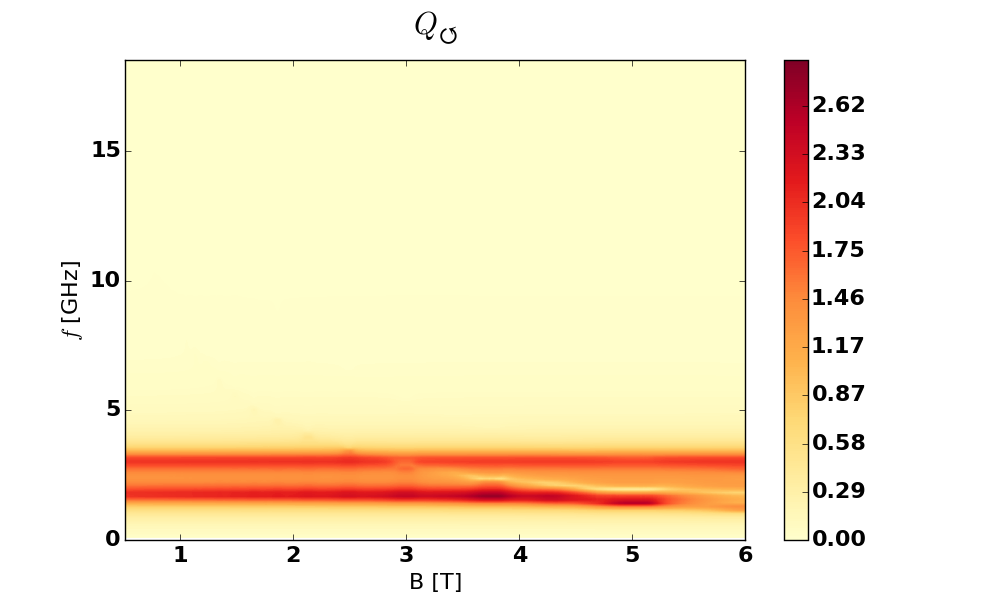}
    \caption{Effects of parasitics on the $Q$-parameters of a three-port circulator. The circulator is matched to the external impedance of $50\Omega$ and the matching is optimized for filling factor $\nu = 8$.
    We used here $C_p=120$fF. The currents carried by the parasitic capacitances and the direct channels are comparable, thus interference effects occur and reversed circulation is strongly enhanced.}
    \label{fig:S-circ-parasitics}
\end{figure}

\begin{figure}[h!]
    \centering{}
    \includegraphics[width=\textwidth]{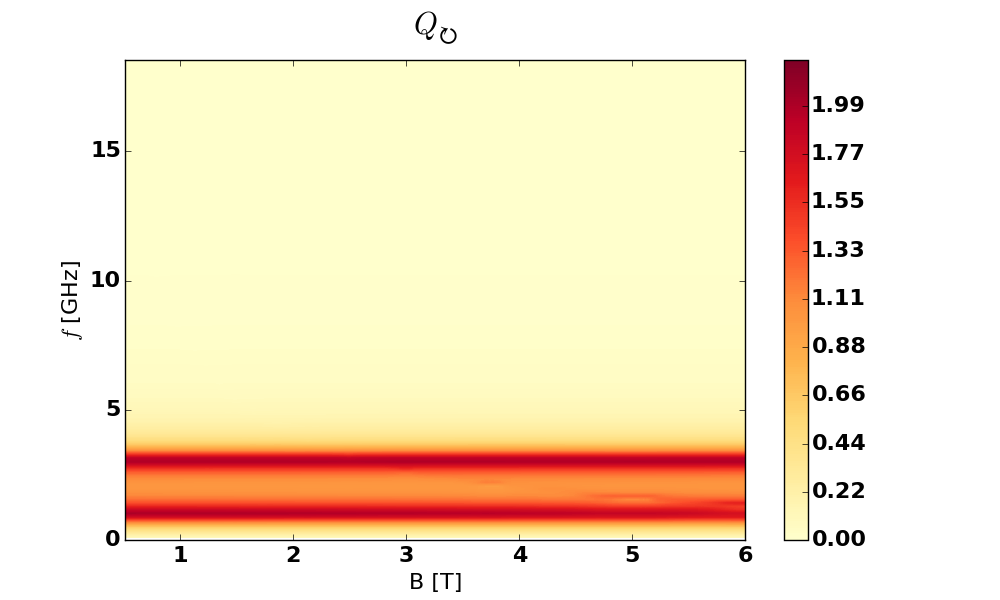} \\
    \includegraphics[width=\textwidth]{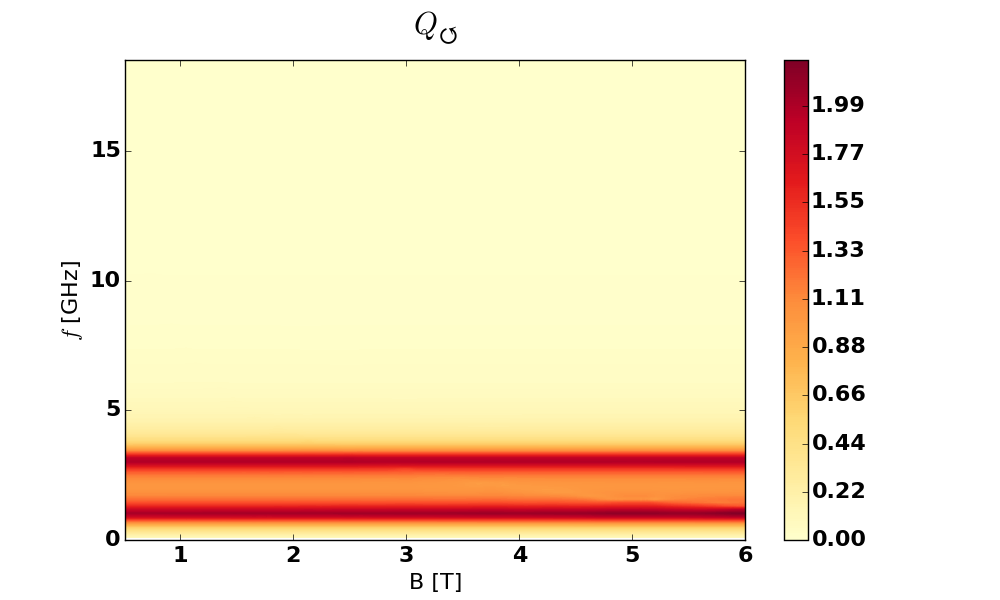}
    \caption{Effects of parasitics on the $Q$-parameters of a three-port circulator. The circulator is matched to the external impedance of $50\Omega$ and the matching is optimized for filling factor $\nu = 8$.
    We used here $C_p=500$fF. As the parasitics are much greater than the direct capacitors, they dominate the response, and circulation is suppressed in both directions.}
    \label{fig:S-circ-strong-parasitics}
\end{figure}

\begin{figure}[h!]
    \centering{}
    \includegraphics[width=0.8\textwidth]{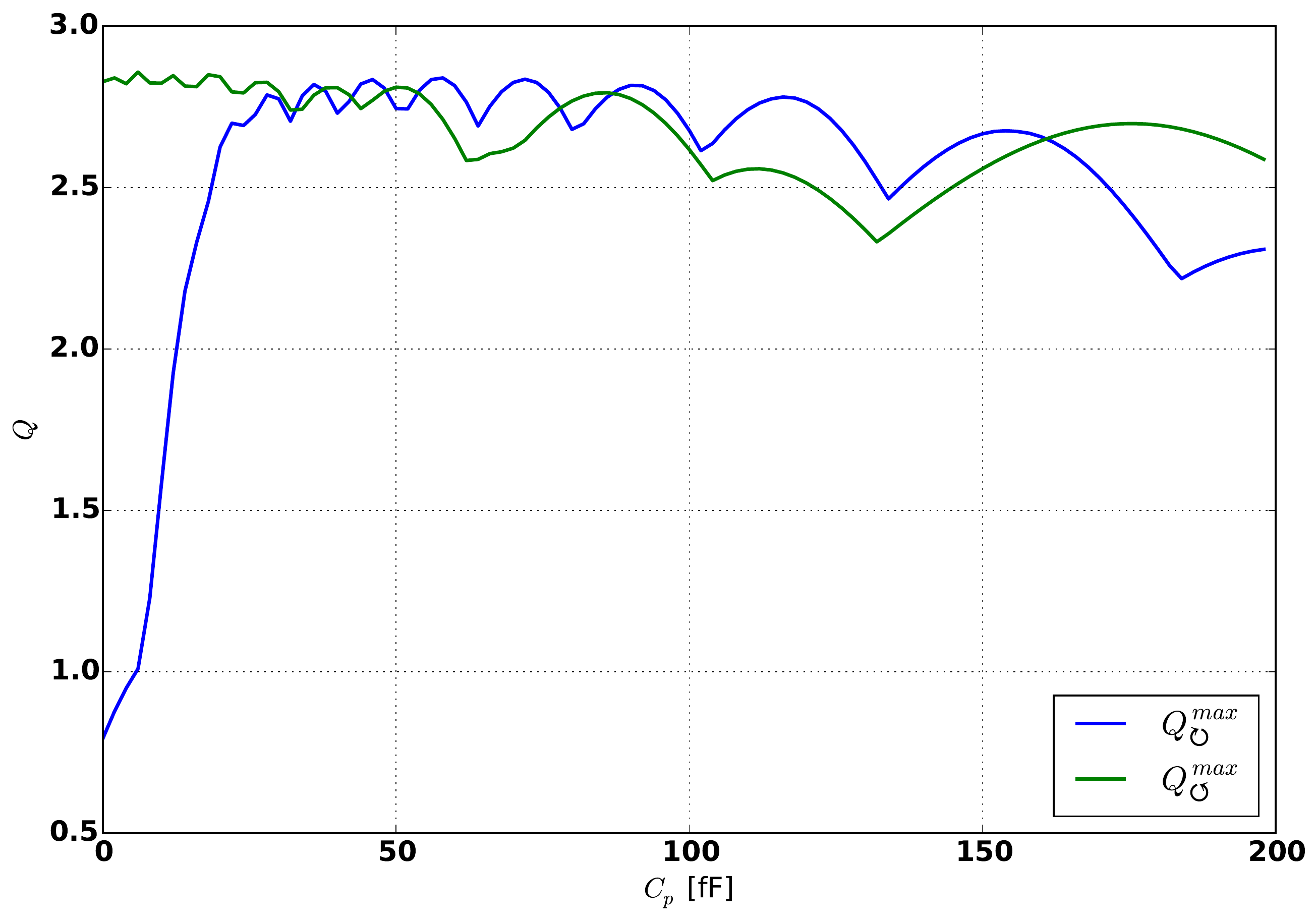}
    \caption{Maximum value of the $Q$-parameters as a function of the parasitic capacitance $C_p$. The circulator is matched to the external impedance of $50\Omega$ and the matching is optimized for filling factor $\nu = 8$. 
As expected, when the currents carried by the parasitic capacitances and the direct channels are comparable, i.e. $C_p\approx 20$fF, circulation in the reversed direction is strongly enhanced. Interestingly, for certain values of $C_p$, e.g. $C_p\approx40$fF, circulation is equally good in both directions; thus, tiny changes in the operating frequency of the device can reverse the circulation direction.}
\label{fig:Q-of-Cp}
\end{figure}

\end{backmatter}
\end{document}